\documentclass[letterpaper]{aa} 
\usepackage{graphicx}
\usepackage{txfonts}
\usepackage{natbib}

\usepackage{graphicx}

\newcommand{\vect}[1]{\ensuremath{{\bf #1}}}

\newcommand{\unit}[1]{\ensuremath{\mathrm{#1}}}

\providecommand{\micron}{\ensuremath{\mu\mathrm{m}}}

\begin{document}
  \title{Measurement of Antenna Surfaces from In- and Out-Of-Focus
    Beam Maps using Astronomical Sources}

\titlerunning{Measurement of Antennas Using Out-of-focus Beam Maps}

  \author{B. Nikolic, R.~E.~Hills \and  J.~S.~Richer }

  \institute{Mullard Radio Astronomy Observatory, Cavendish
  Laboratory, Cambridge CB3 OHE, UK }

  \date{ }

  \abstract{We present a technique for the accurate estimation of
  large-scale errors in an antenna surface using astronomical
  sources and detectors.   The technique requires several out-of-focus
  images of a compact source and the signal-to-noise ratio needs to be good
  but not unreasonably high.
  For a given pattern of surface errors, the expected form of such images can be
  calculated directly.  We show that it is possible to solve the inverse problem
  of finding the surface errors from the images in a stable manner using standard
  numerical techniques.  To do this we describe the surface error as a linear
  combination of a suitable set of basis functions (we use Zernike polynomials).
  We present simulations illustrating the
  technique and in particular we investigate the effects of receiver
  noise and pointing errors.   Measurements of the 15-m James Clerk
  Maxwell telescope made using this technique are presented as an
  example. The key result is that good measurements of errors on large
  spatial scales can be obtained if the input images have a
  signal-to-noise ratio of order 100 or more. The important advantage of
  this technique over transmitter-based holography is that
  it allows measurements at arbitrary elevation angles, so allowing one
  to characterise the large scale deformations in an antenna as a
  function of elevation. 

 }

\maketitle

\section{Introduction}

Measuring the shape and deformations of radio antennas using microwaves
(often called holography) is useful both for correcting the surface errors
directly, and for testing theoretical models of the behaviour of telescope
structures.  Both of those are in turn invaluable for the
characterisation of current and future generations of high-precision
10-20\,m class antennas which are required to have high gain at frequencies
up to and beyond 1\,THz, and of course for larger antennas operating at
lower frequencies.  Maximising the surface accuracy of existing
and new antennas is of increasing scientific importance for two
reasons: it increases the gain (particularly at higher frequencies),
so allowing detection of fainter compact sources, and it reduces the
amount of power in the `error beam', so allowing imaging of extended sources
with higher dynamic range.

In the phase-retrieval approach to millimetre-wave holography, only
the power pattern of the antenna is measured, usually at two or more
different focus settings. The phase of the signal in the aperture is
later recovered by numerical processing \citep{Morris1985,Anderson1985}.  
This technique has been applied with considerable success on a number
of large antennas, but usually only with artificial sources,
i.e. transmitters on the ground or on spacecraft
\citep[e.g.,][]{1993A&A...274..975F}. This is typically possible only
at a small range of elevations: for ground based beacons, for example,
measurements are essentially only possible with the telescope pointed
within ten degrees or so of the horizon, and this does not allow
measurement of gravitation deformations of the surface.

This paper describes the development of a technique for measuring
surface errors with moderate spatial resolution by observing
astronomical sources, using existing astronomical receivers on the
telescopes.  The new approach uses numerical fitting of a
parameterised description of the surface errors, and of the amplitude
of the receiver's illumination pattern.  The technique is flexible.
It can be adapted straightforwardly to various different observational
techniques, including total power observations, and the various
differencing schemes which involve movement of the secondary reflector
(`beam switching') and/or primary reflector (`nodding').  Extended
sources, such as planets, can also be accommodated.  In future, the
increasing availability of astronomical array detectors will make the
mapping process particularly quick and efficient.

\section{The technique}

If both the amplitude and the phase of the far-field beam pattern of a
telescope are measured, a simple Fourier inversion will give the
aperture function and thus the deformations of the telescope. This is
the conventional \emph{with-phase} holography which is widely used for
measurement of radio-telescopes either when they already operate as
parts of interferometers \citep{1977MNRAS.178..539S} or when equipped
with an auxiliary phase-reference antenna
\citep[e.g.,][]{1988A&A...203..399M}.  Measuring just the power
pattern of the in-focus beam under-determines the aperture
function---for example, reversing the sign of surface errors produces
no change in the power pattern. This degeneracy can be broken, and,
additionally, partially independent data sets obtained, by introducing
a known phase change across the aperture, for example by defocussing
the telescope by a known amount.  It is also true that small errors in
the surface of the reflector have a much greater effect on defocussed
images than on the in-focus one.  This is the main reason why the
method described here works well with only moderately high signal to
noise ratio.

Due to the non-linearity of the measurement process, a straightforward
inversion from the observed beam power patterns to the aperture plane
phase distribution is not possible. Rather, the aperture needs to be
parameterised in terms of a suitable set of basis functions and a
numerical fitting procedure employed.

We parameterise the aperture phase distribution -- which is determined
by the combination of surface errors, phase response of the receiver
and any mis-collimation of the telescope -- in terms of a fixed number
of coefficients of Zernike circle polynomial functions, i.e.:
\begin{equation}
\phi(x,y) = \sum_{n=1}^{n_{\rm max}} \quad \sum_{l=-n, -n+2, ... ,n} a_{n,l}
Z_{n,l}(x,y), 
\end{equation}
where $\phi(x,y)$ is the aperture phase distribution, $a_{n,l}$ are
the coefficients to be determined and $Z_{n,l}(x,y)$ are the Zernike
circle polynomials labeled with to their radial ($n$) and angular
($l$) orders (see, for example, \citealt{BORNWOLF} for a
definition). For each radial order $n$ there the are $n+1$ possible
angular orders, and so the parameter $n_{max}$ determines the total
number of coefficients [$(n+1)(n+2)/2$] which are to be found by
fitting.

For example, if the maximum radial order is six (approximately the
value typically used in practice, see later and
\citealt{OOFNikolic06p2}), there are a total of 28 coefficients to be
found.  The smallest scales present in the highest order (n=6)
functions correspond to scales of roughly one fifth the dish diameter.

This choice of parametrisation in terms of Zernike polynomials has
several advantages, in particular that they are orthogonal on the unit
circle. This strict orthogonality breaks down when applied to
practical measurements of radio antennas because the illumination is
not uniform across the aperture, due to the edge taper and central
blockage.  Nonetheless, approximate orthogonality is maintained.
Further, by restricting the highest order of Zernike polynomial used,
the results can be made less sensitive to the poorly-constrained
small-scale errors on the surface. Finally, some of the low-order
Zernike polynomials correspond closely to aberrations which are likely
to be present in the optics. For example, a misalignment of the
secondary reflector in a Cassegrain telescope away from the optical
axis gives rise to a aberration which is closely approximated by the
$n=3$, $l=\pm1$ Zernike polynomials.

To model the far-field antenna power pattern, besides the phase of the
aperture function, it is also necessary to know its amplitude [often
referred to as the illumination, $I(x,y)$, of the primary reflector].
This is determined in practice by the properties of the receiving
system which couples detectors to the primary antenna surface.  When
these are known with sufficient accuracy, the amplitude distribution
of the aperture function can be determined a priori and taken as a
given in the model.  We have however found that it is generally better
to model the amplitude as a function of a small number of free
parameters and solve for these simultaneously with the coefficients
$a_{n,l}$ that determine the phase. We use a two-dimensional Gaussian
as a model for the amplitude with free parameters describing its
width, offset with respect to the axis, etc, i.e.:
\begin{equation}
  I(x,y)=I_{0}\exp\left[-\frac{(x-x_{0})^{2} +
  (y-y_{0})^{2}}{R^{2}}\sigma_{r}\right],
\end{equation}
where $R$ is the radius of the primary reflector, $(x_{0}, y_{0})$
define the centre of the illumination on the primary reflector and
$\sigma_{r}$ is the illumination taper.

If the best-fitting model amplitude is not
in agreement with the expected amplitude distribution, then this may
indicate a problem in the way the data have been processed or a
problem in the receiving system.

\subsection{Phase change due to defocus}

For a Cassegrain telescope, in the ray-tracing approximation, the
extra path $\delta$ due to a radial defocus of the telescope by a
distance $d_{Z}$ is given by:
\begin{equation}
  \delta(x,y; d_{Z}) = d_{Z} \left( \frac{1 - a^{2}}{1+a^{2}} + \frac{1 -
  b^{2}}{1+b^{2}}  \right)
\label{eq:dzphase}
\end{equation}
where $a = r / (2f) $, $b = r / (2F) $, $r=\sqrt{( x^{2}+y^{2})}$ is the radius in the
aperture plane from the optical axis of the telescope, $f$ is the
focal length of the primary reflector, $F$ is the effective total
focal length of the telescope at the Cassegrain focus, and a positive
value of $d_{Z}$ corresponds to a movement of the secondary reflector
away from the primary.

Putting all of the above together, the aperture function $A(x,y)$ may
be written as:
\begin{equation}
  A(x,y)= \Theta\left(R^{2} - x^{2}- y^{2}\right)
  I(x,y)\exp\left[\phi(x,y)+\delta(x,y;d_{Z})\right]
\label{eq:apfn}
\end{equation}
where $\Theta$ is the Heaviside step function which described the
truncation of the aperture function at the edge of the primary
reflector.

\subsection{The fitting procedure}

The parametrised form of the aperture function
(Equation~\ref{eq:apfn}) may be easily computed on a convenient
rectangular grid making it straightforward to calculate the
corresponding power beam pattern using the fast Fourier transform.
The expected out-of-focus beam patterns are computed by adding to the
aperture function, prior to the Fourier transform, the additional
phase change due to the defocus as given by Equation~\ref{eq:dzphase}
for the case of a Cassegrain telescope.  If the measurement was not of
the true antenna power pattern --- as may be case if a differencing
technique is used, or if the source employed was not sufficiently
point-like --- this can be taken into account by an appropriate
convolution of the model power-pattern, in either the sky or the
Fourier domain.

In the procedure we use, the model power beam pattern is compared to
observed data (labeled $\vect{D}$) by interpolating the model produced
using the fast Fourier Transform to the position of each observed data
point (producing the model vector $\vect{y}$), eliminating the need
for the observed data to be on a regular grid or for it to be
re-gridded in any way. This is desirable since interpolation and
regridding of the observed data generally causes a loss of information
whereas interpolating the model does not.  The computational burden of
this step is reduced by storing the interpolation coefficients between
iterations.

The agreement between a model and observed data is characterised by
the vector of residuals, ${r}_{i}$, weighted by an estimate of
the measurement error, $\sigma_{i}$, at each data point; i.e.,:
\begin{equation}
{r}_{i} = \frac{D_{i} - y_{i}}{\sigma_{i}}.
\end{equation}
In the simplest case, the measurement errors will be dominated by
thermal noise characterised by the system temperature, the bandwidth
used and the integration time. In practice, other factors can
significantly contribute to measurement errors. These may include
imperfect removal of a time-varying atmospheric emission, variation in
the gain of the receiver, source noise (in the case of observations of
very strong narrow line sources, such as SiO masers), pointing errors
of the telescope and radio ``seeing'' (direction of arrival
fluctuations due to the atmosphere).

The best fitting telescope surface is found by minimising the
magnitude of the error-weighted residual vector, i.e., $|r|$. At this
point a ``prior'' on the surface deformations, or equivalently, on the
phase of the aperture function, may be added to the minimisation
process as an extra residual.  For example one can add to this
quantity the value of the root-mean-square of the surface error in the
model, multiplied by a constant, do the minimisation and then inspect
the resulting match between the model beams and the data as the value
of this constant is varied.  A large value of the constant favours
solutions with small surface errors, so by gradually reducing it one
can choose the most conservative solution which provides an adequate
fit.  This is helpful if, for example, one is using the measurements
to make adjustments of the surface.

Our initial investigations employed the downhill-simplex minimiser,
and we found that for small to moderate phase errors (i.e. less than a
whole turn of phase peak-to-peak) the posterior distribution is in
general convex and with a well defined global maximum.  This means
that it should be sufficient to do a simple search for the maximum of
the posterior distribution rather than characterise it in full. We
therefore routinely employ the relatively efficient
Levenberg-Marquardt algorithm \citep{PDASTAR}.

\section{Simulations}

The feasibility and accuracy, and the optimum observational approach,
of the technique just described has been investigated using numerical
simulations. These simulations have been carried out for a Cassegrain
antenna with parameters similar to the proposed Atacama Large
Millimetre Array (ALMA) antennas, i.e., with a primary reflector of
12\,m diameter, effective focal lengths of 4.8\,m and 96\,m at primary
and Cassegrain foci respectively, and a Gaussian illumination with a
12\,dB edge taper. The main results can be straightforwardly
reinterpreted for different antennas. The wavelength of simulated
observations was 1\,mm.

The approach taken was to calculate theoretical beam maps, both in-
and out-of-focus, for a perfect dish and to add to these beam-maps
simulated noise and instrumental effects. These were then used as
input to the fitting procedures. 

The accuracy of the best-fitting surface is reported as $\epsilon$,
the illumination-weighted root-mean-square of the phase of the
aperture function.  Since $\epsilon$ scales with wavelength, we
express it in the of units of radians of phase, which at a given
observing frequency are easily converted to a physical length. For
example, at a typical observing frequency of 230\,\unit{GHz} and
assuming normal incidence (i.e., a large $f$-ratio aperture), a
wavefront error of 0.1\,radian corresponds to a deformation of the
surface of 10.4\,\micron.

The fractional decrease of the on-axis antenna gain due to these
errors is given approximately by $\epsilon^{2}$ \citep{RuzeEff}, so
the results of the simulations can be interpreted as follows. If the
telescope is dominated by large scale errors, then after resetting a
surface with an out-of-focus holography solution with $\epsilon=0.1$,
then the efficiency of the telescope at the frequency at which the
out-of-focus beam maps were made would be $\approx 99\,\unit{\%}$.
The efficiency at three times that frequency will, however, be only
$\approx 91\,\unit{\%}$.

The results of these simulations are strictly only applicable to
antennas with surface errors much smaller than the observing
wavelength. For antennas with significant deformations the errors on
the measured surface may be expected to be somewhat larger. If,
however, an iterative approach to correcting and re-measuring the
antenna is taken, then it should be possible to approach the
accuracies which are given below.

\subsection{Effect of Noise in Input Maps}

We first consider the relationship between errors on the derived
surface, the signal to noise ratio and the number of coefficients of
Zernike polynomials being fitted for. A number of noisy realisations
($\vect{D}_{\rm s}'$) of the theoretical beams ($\vect{D}_{\rm s}$)
were derived using:
\begin{equation}
\vect{D}_{\rm s}'=   \vect{D}_{\rm s} (1 + \vect{m}) + \vect{a}
\end{equation}
where $\vect{m}$ and $\vect{a}$ are random Gaussian variables with
zero mean such that $\langle \vect{m}^{2} \rangle^{\frac{1}{2}} = m$
and $\langle \vect{a}^{2} \rangle^{\frac{1}{2}} = a$. The
multiplicative noise is then simply determined by $m$, while we set
$a$ so that the peak signal-to-noise is given by
$\max\left(\vect{D}_{\rm s}\right) / a$.  

\begin{figure}
\includegraphics[clip,width=\linewidth]{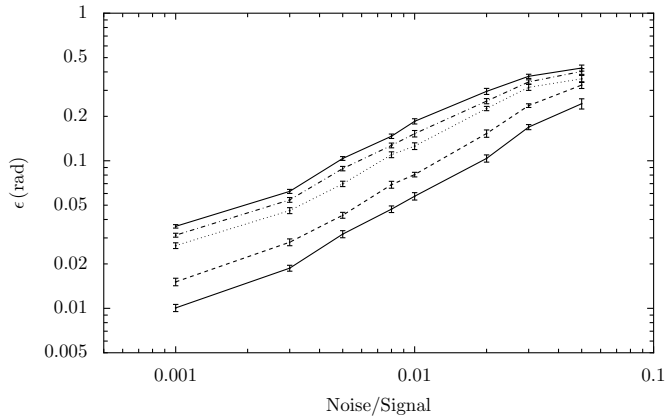}
\caption{The weighted root-mean-square error ($\epsilon$, in the units
  of radians of phase) of the best-fitting aperture phase as a
  function of peak signal to noise.  The five traces correspond,
  bottom to top, to derived errors when fitting for coefficients of
  Zernike polynomials up to 4th, 5th, 6th, 7th and 8th radial
  order. The out-of-focus maps were simulated at $\pm 2\,\unit{mm}$
  defocus.}
\label{fig:additivesn}
\end{figure}

The effect of thermal noise is illustrated in Figure
\ref{fig:additivesn}, which shows the surface errors from a number of
simulated data sets of different signal-to-noise ratios, all with
multiplicative noise set to $m=0.05$ and consisting of three
Nyquist-sampled maps, one in-focus and one at -2\,mm and +2\,mm
defocus each. Derived surfaces were calculated separately for
parameterisations with maximum radial orders of Zernike polynomials
ranging from 4 to 8, thereby quantifying the increase in measurement
error with increasing spatial resolution.

\begin{figure}
\includegraphics[clip,width=\linewidth]{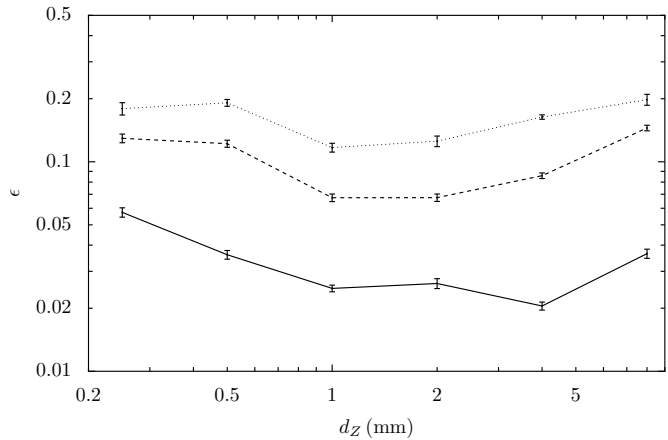}
\caption{The weighted root-mean-square error ($\epsilon$, units
  radians of phase) of the best-fitting aperture phase as a function
  of the magnitude of defocus used in the out-of-focus maps for three
  noise to signal ratios (bottom to top): 0.001, 0.005, and 0.01. The
  maximum radial order of Zernike polynomials used in the fit was 6
  and a multiplicative noise of 5 percent was added to the maps as before.}
\label{fig:dzstudy}
\end{figure}

The relationship between the degree of defocus used for the
out-of-focus maps and the accuracy with which the aperture phase
distribution can be recovered is illustrated in
Figure~\ref{fig:dzstudy}. Since the area over which power is spread
increases with increasing defocus, it may be expected that the
accuracy is a function of the signal to noise in the maps. For this
reason, simulations with three signal to noise ratios are shown in the
figure, indicating that at better signal to noise ratios, a somewhat
larger defocus produces optimal results.  For this case, however, with
a significant multiplicative noise level, the magnitude of the defocus
is not very critical.

\subsection{Effect of Pointing Errors}

We next consider the extent to which pointing jitter degrades the
accuracy of the inferred aperture phase. Such jitter may be the result
of servo tracking errors, radio seeing or rapidly varying deformations
of the telescope structure. The simplest way to model such errors in
the pointing is by assuming that each sample in the observed maps is
displaced by a random distance $\delta \theta, \delta\phi$ from its
true position, with these displacement being Gaussian random variables
with variances of $\Delta^2$ and there being no correlation between
samples.

\begin{figure}
\includegraphics[clip,width=\linewidth]{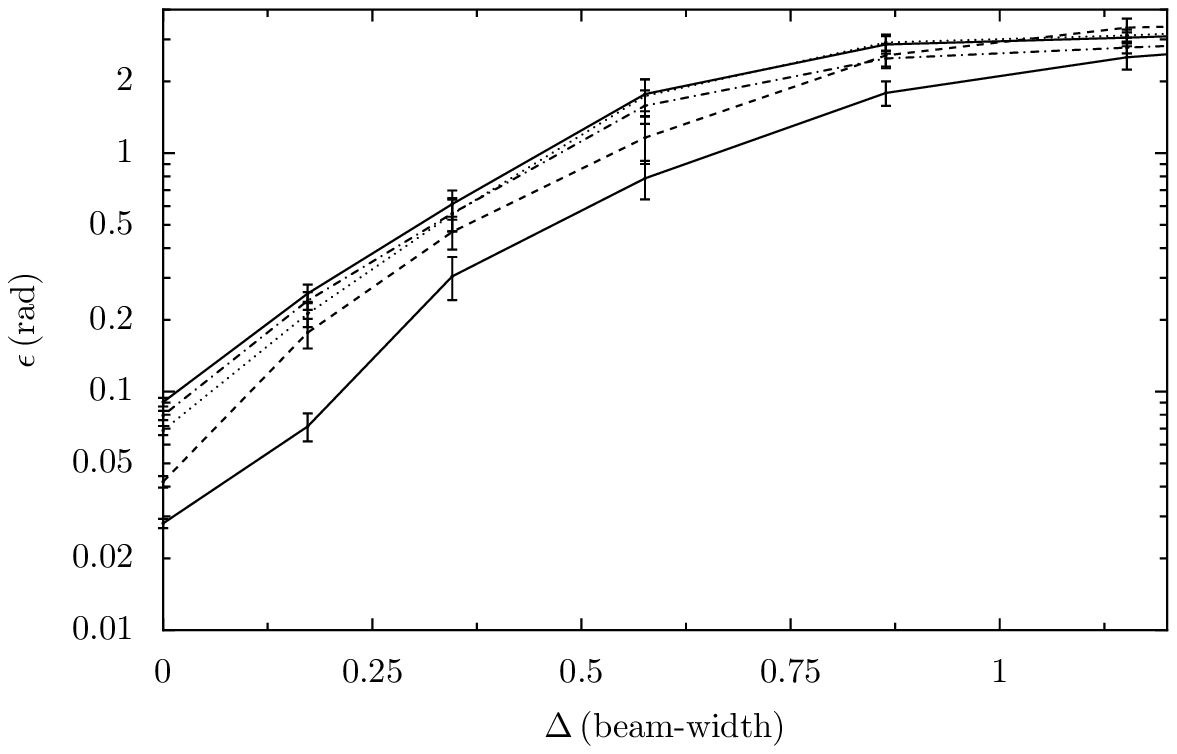}
\caption{Calculated surface error ($\epsilon$, units radians of phase)
  as a function of $\Delta$, the root-mean-square of pointing errors,
  which are assumed to be uncorrelated between the samples. The traces
  correspond to, from bottom to top, to simulations with maximum
  Zernike radial orders of 4, 5, 6, 7 and 8. The thermal signal to
  noise ratio is 200 to 1 with a five per-cent multiplicative noise
  as before.}
\label{fig:jitter}
\end{figure}

Figure~\ref{fig:jitter} shows the effect of such pointing errors in a
set of simulated maps with a peak signal to noise ratio of 200 to 1
and 5 percent multiplicative noise. It can be seen that the accuracy of
the technique rapidly gets worse as pointing errors become
significant. For example, RMS pointing jitter of 0.2 beam-widths
increases the error in the derived surface by more then a factor of
two when compared to the no-pointing jitter case.

We have also explored the impact on accuracy of the technique of using
resolved sources for the beam maps. We find that for sources with
sharp edges the accuracy is not much affected by their size.  This
means that objects like Venus and Mars are good sources for such
measurements, but that Jupiter is somewhat less so because of limb
darkening.  We have even succeded in using Saturn when this was the
only source available, but this did require a good model for the
source.

\section{A sample measurement at the James Clerk Maxwell Telescope}

The James Clerk Maxwell Telescope is a 15-m submillimetre telescope
with conventional Cassegrain optics, covering the atmospheric
transmission bands from 150\,GHz to 1.5\,THz (2\,mm to 200\,$\mu$m
wavelength): its target surface accuracy is around 22\,$\mu$m. We here
present some sample out-of-focus holography measurements taken at the
JCMT as an illustration of the technique. 

\begin{figure*}[t]
\includegraphics[clip,angle=-90,width=0.3\linewidth]{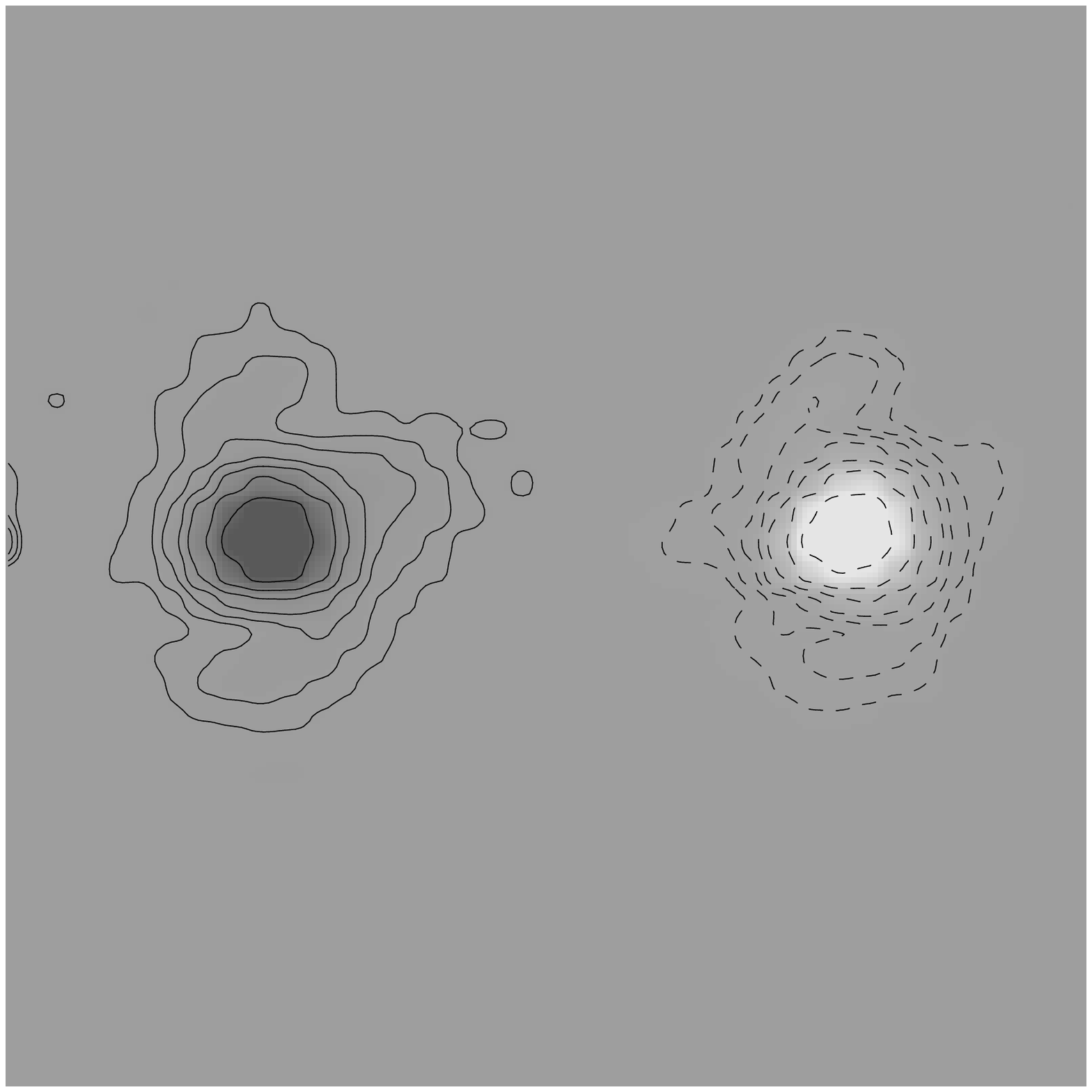}
\includegraphics[clip,angle=-90,width=0.3\linewidth]{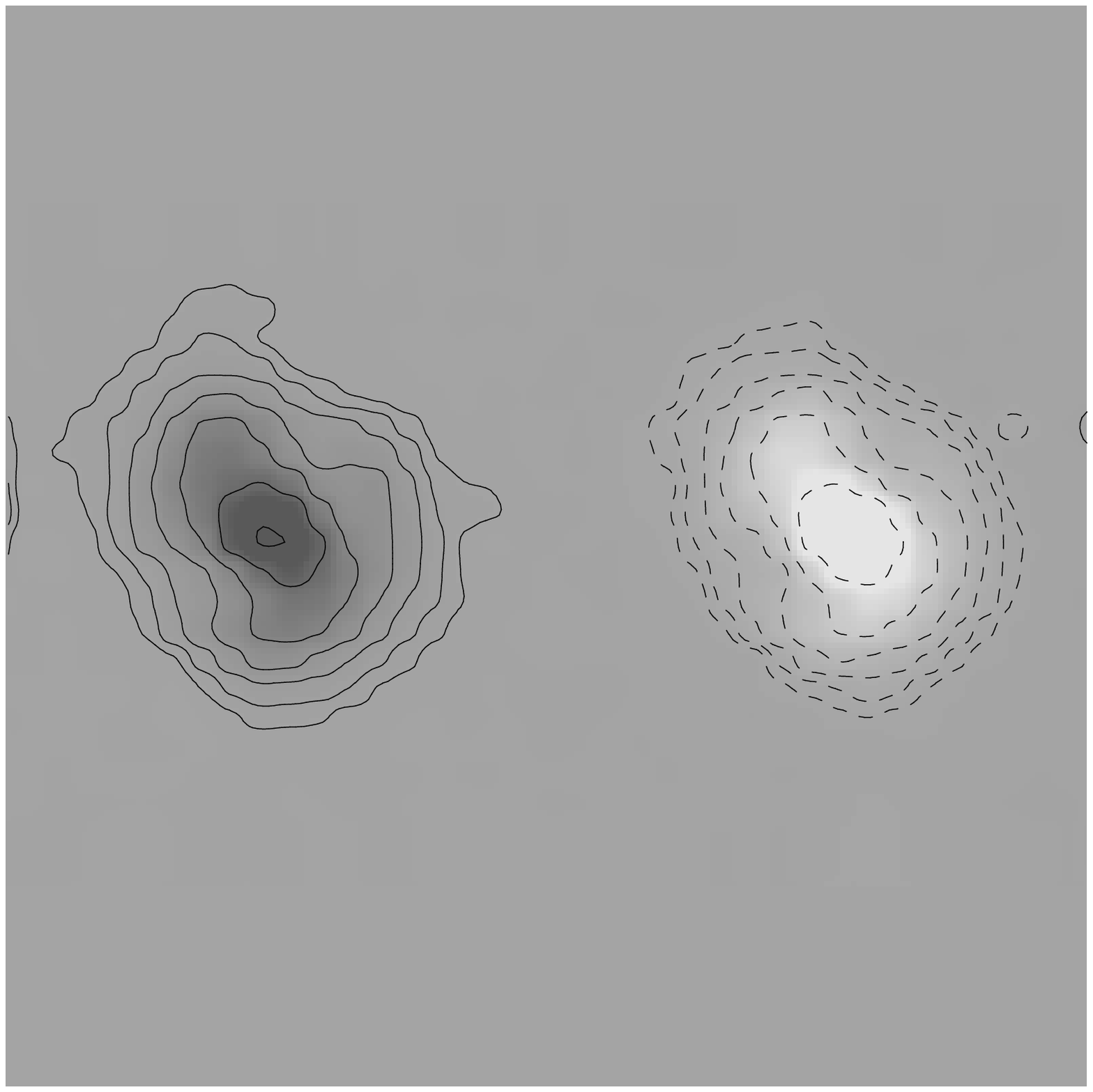}
\includegraphics[clip,angle=-90,width=0.3\linewidth]{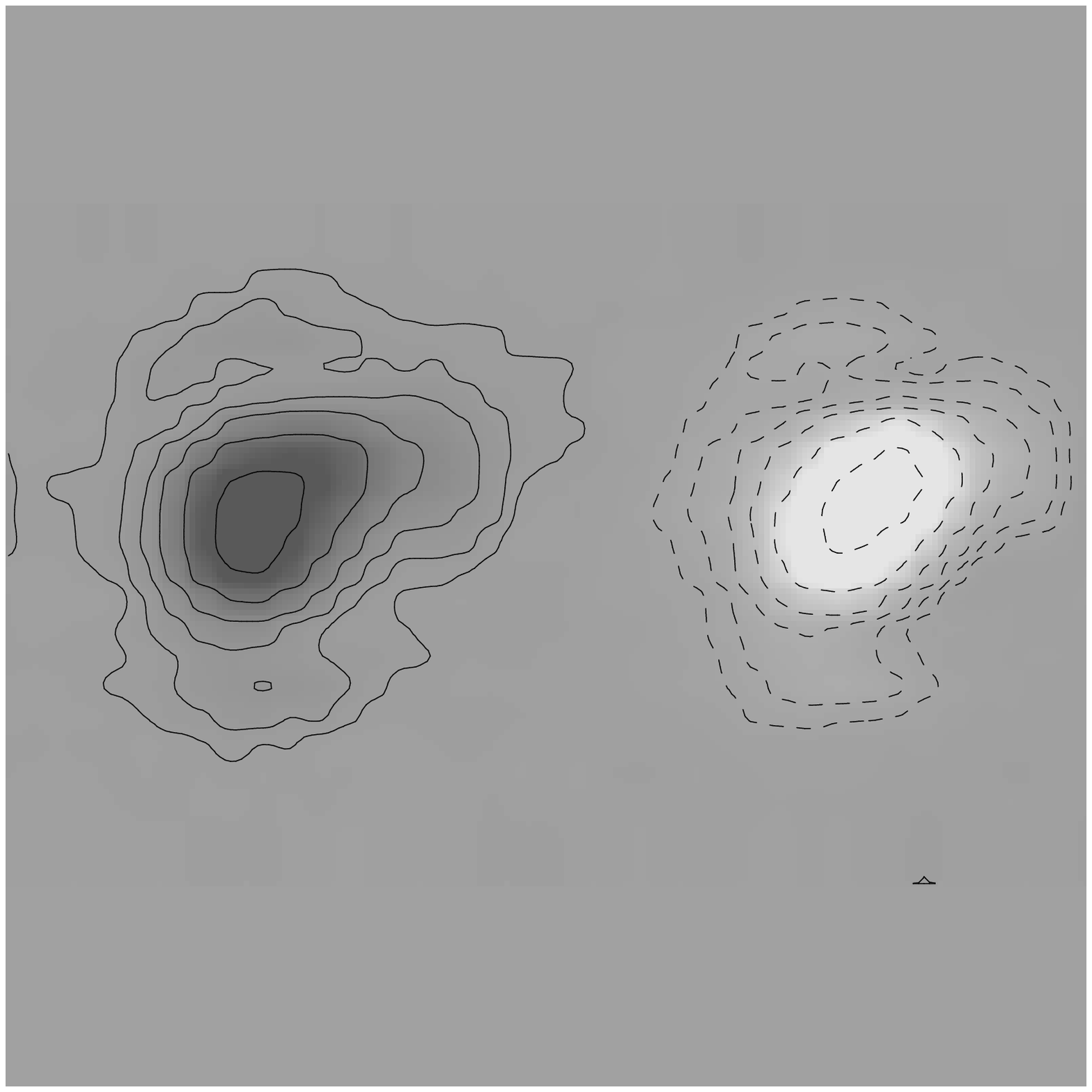}

\includegraphics[clip,angle=-90,width=0.3\linewidth]{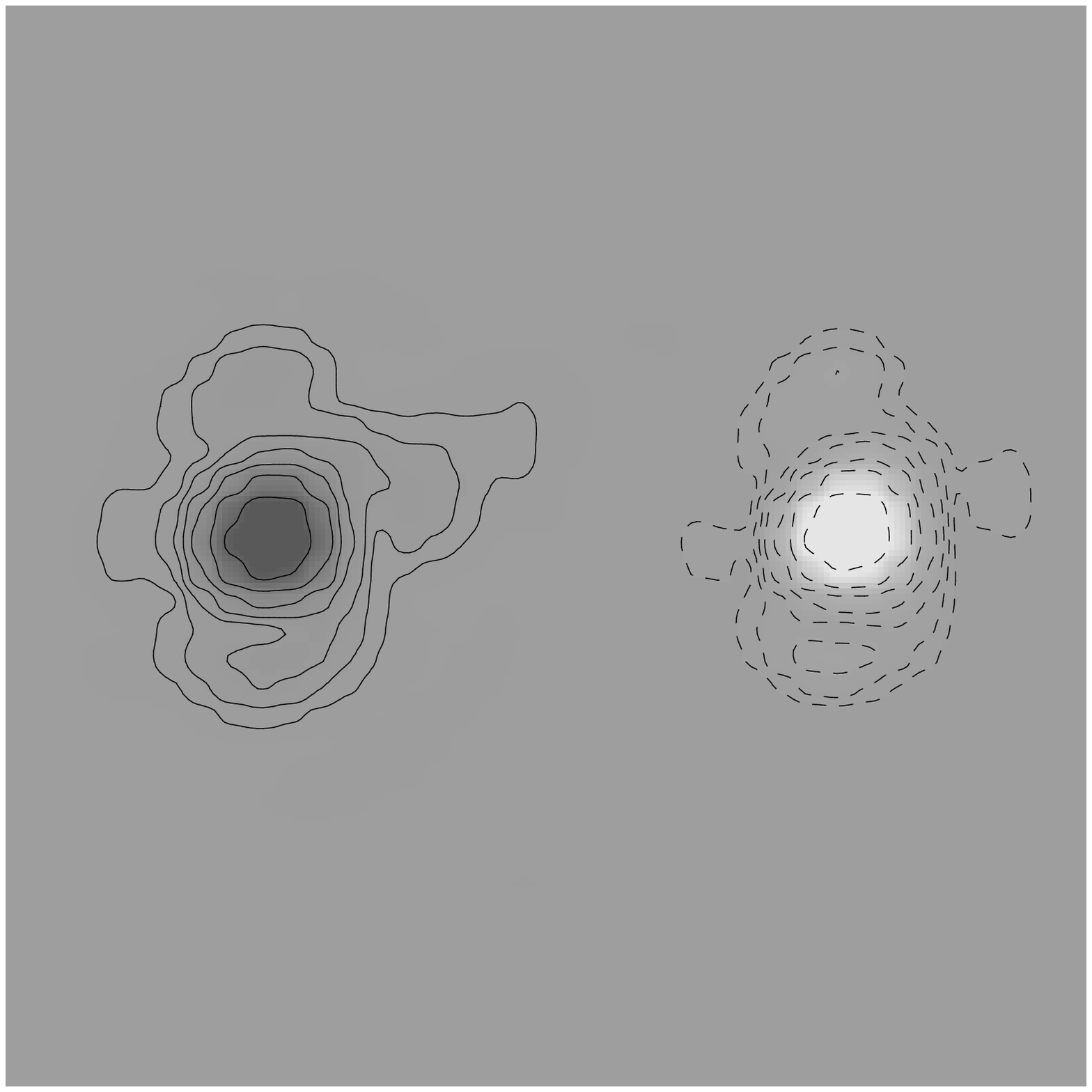}
\includegraphics[clip,angle=-90,width=0.3\linewidth]{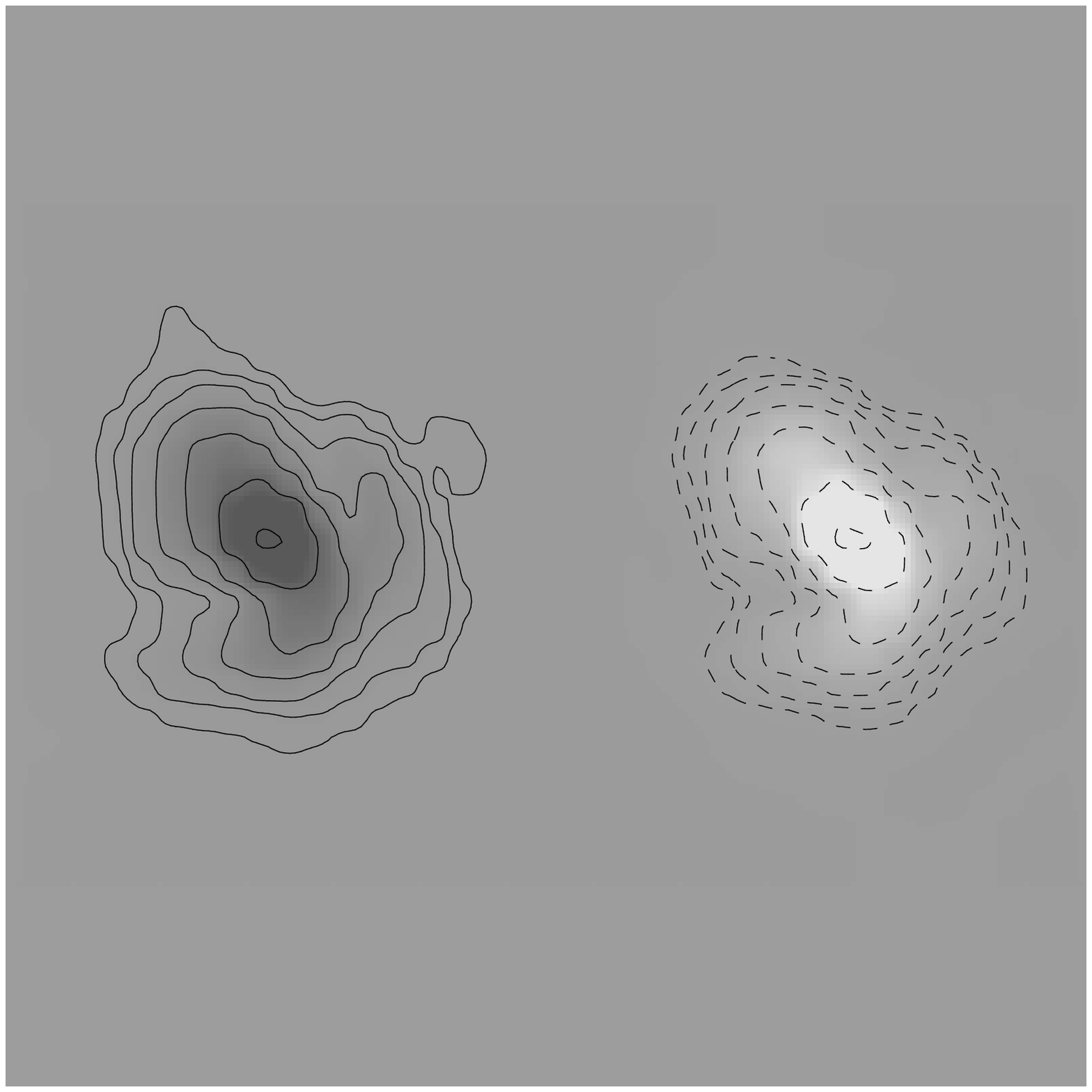}
\includegraphics[clip,angle=-90,width=0.3\linewidth]{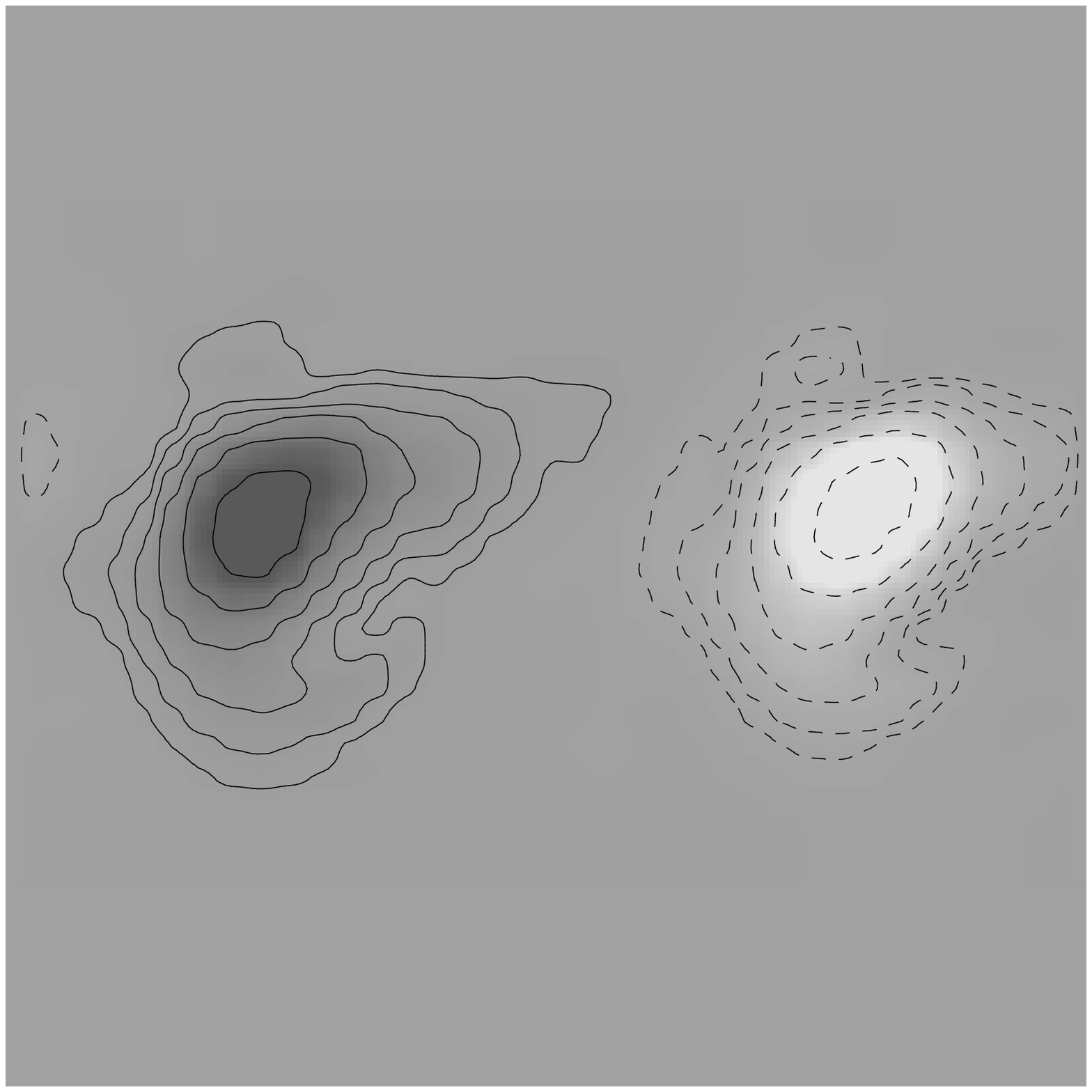}

\caption{Comparison of observed in- and out-of-focus beam maps (top
  row, left to right: in-focus, +2\,mm and -2\,mm out-of-focus) and
  best-fitting model beam maps (using up to $6^{\rm th}$ order Zernike
  polynomials, bottom row, same order as top row). Contours are at
  intervals $2^{-n}$ (thus, the lowest contours are at -21\,dB in the
  two leftmost maps and at -18\,dB in the rightmost map which has one
  contour less). It is seen that the action of the chopping secondary,
  followed by synchronous detection, is to produce two beams -- one
  positive and one negative.}
\label{fig:dzmaps}
\end{figure*} 

These observations were made on Venus with a heterodyne receiver tuned
to a frequency of 232.5\,GHz, and utilising a chopper (with an 160
arcsecond throw) for atmospheric rejection.  The apparent diameter of
Venus at the time of observations was 12 arcseconds, smaller than the
main beam of the telescope (which at this frequency has the full-width
half-maximum of 20 arcseconds), but large enough to require taking
into account in the analysis. In total five maps were taken, although
here we concentrate only on the in-focus and $\pm 2\,\unit{mm}$
out-of-focus maps. All of the maps were 320 arcseconds long in the
azimuth direction and 160 arcseconds long in the elevation direction
and were sampled at 8 arcsecond intervals. They are shown in the top
row of Figure~\ref{fig:dzmaps}.

The large chop throw used in these observations means that it is
necessary to take into account the aberrations introduced by the tilt
of the secondary reflector which is required to deflect the beam. The
phase change in the aperture plane corresponding to this aberration
has been calculated using a ray-tracing software package: the dominant
term is, as usually is the case, coma, with a magnitude of 0.072
radians at dish edge for a throw of 80 arcseconds on the sky (i.e.,
the deflection from the centre beam for the throw used).

\begin{figure}
\includegraphics[clip,angle=-90,width=\columnwidth]{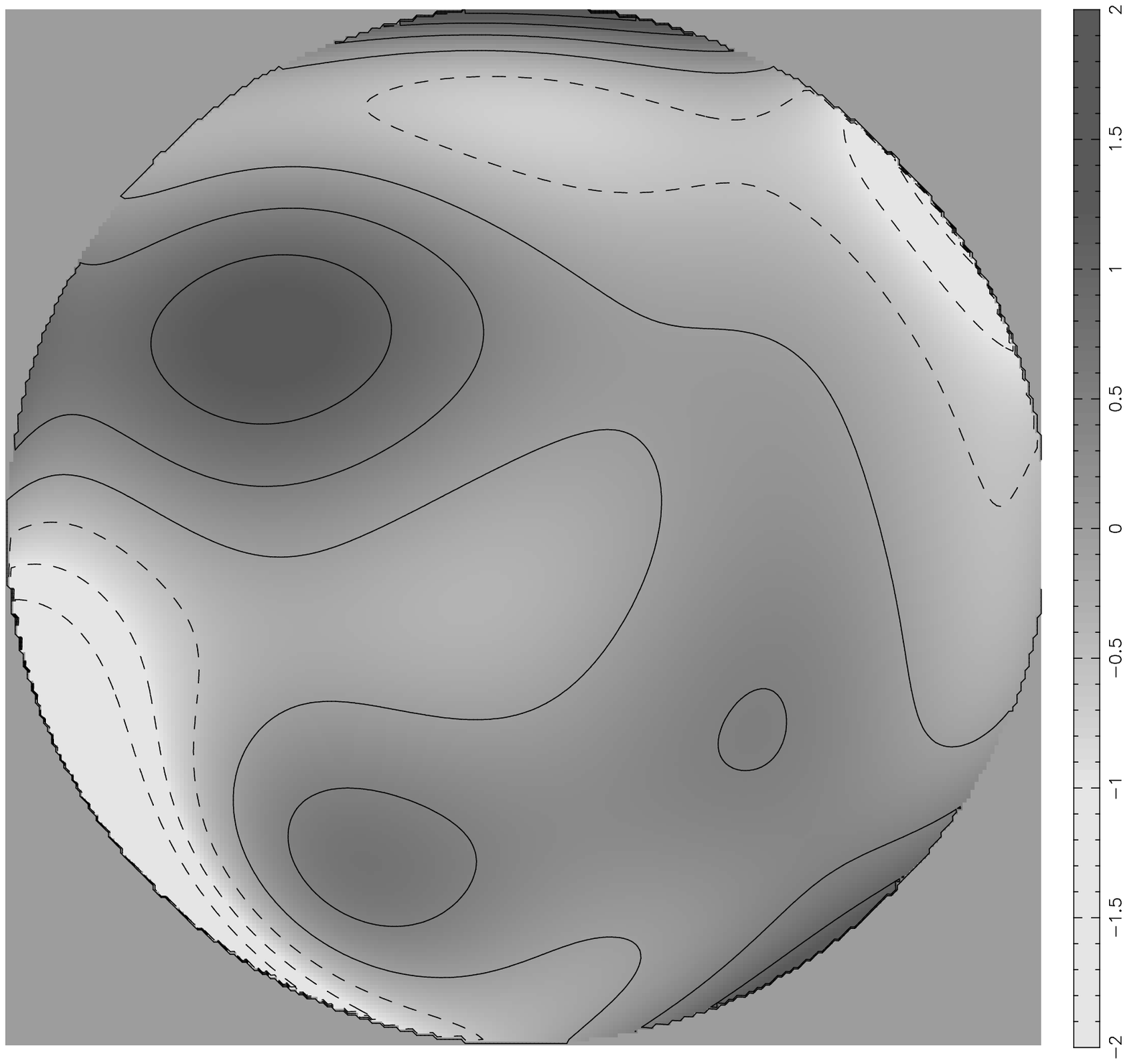}
\caption{The inferred aperture-plane phase distribution (in units of
  radians of phase) from the best-fitting model (using up to $6^{\rm
  th}$ order Zernike polynomials) to data shown in
  Figure~\ref{fig:dzmaps}. Contours are at fixed intervals of 0.5
  radians (corresponding to 51\,\micron). Positive values represent
  bumps while negative values represent dips on the surface of the
  dish.}
\label{fig:jcmtphase}
\end{figure}

The best-fitting surface map is shown if Figure~\ref{fig:jcmtphase}
and the corresponding model beams in the bottom row of
Figure~\ref{fig:dzmaps}. These measurements were taken at a time when
there was a known fault in the dish which resulted in an observed bump
in one of the quadrants of the dish. They \emph{do not} represent the
current status of the JCMT surface.

\cite{OOFNikolic06p2} present a detailed study of the Green Bank
Telescope (a large millimetre-wave telescope), showing how
gravitational and thermal effects can be measured and corrected using
the technique. They also present a detailed practical verification of
the technique.

\section{Discussion and conclusions}

A variation of the phase-retrieval approach to holography has been
presented with the surface parametrised in terms of a basis set of
continuous function in the aperture plane and with optimised defocus
values. Our simulations indicate that relatively moderate dynamic
range (approximately 200 to 1) is required for these
measurements. This means that this technique is feasible with
astronomical sources and normal astronomical receivers.

The technique can only recover large-scale structure of the surface
and probably is not suitable for setting individual panels.
The possibility, of using astronomical sources and
receivers, however, gives several advantages:
\begin{enumerate}
  \item Measurements of the surface can be performed at a wide range
  of elevations.
  
  \item The composite phase response of the entire astronomical
  observing system can be determined, i.e. the effects of
  deformations of the primary and secondary reflectors and the
  response of the feed are all included.
  
  \item No investment in hardware or physical changes to the telescope
  are required.

\end{enumerate}
The combination of first two points above means that the technique
should be highly suitable for measurement of gravitation deformations
of the telescope surface.

Our simulations and current experience have indicated some
drawbacks. One of these is the effect of typically highly tapered
astronomical feeds: this significantly reduces the constraints on the
outer parts of the dish, although, equivalently, these outer parts
contribute less to the effective collecting area of the telescope and
therefore do not need to be determined as accurately. The second is
that pointing/tracking errors can significantly degrade the accuracy
of the technique.

\begin{acknowledgements}

We would like to thank Anthony Lasenby for his suggestions which
contributed to several aspects of the development of this technique.

The James Clerk Maxwell Telescope is operated by The Joint Astronomy 
Centre on behalf of the Particle Physics and Astronomy Research Council 
of the United Kingdom, the Netherlands Organisation for Scientific 
Research, and the National Research Council of Canada.

\end{acknowledgements}

\bibliographystyle{aa} \bibliography{../oofbib}

\end{document}